\documentclass[preprint]{aastex}
\usepackage{graphics}
\usepackage{amssymb,amsmath}
\usepackage{subfig}

\slugcomment{{\bf Accepted by the Astrophysical Journal}}

\begin{document}

\title{Dense, Parsec-Scale Clumps near the Great Annihilator}

\author{Edmund Hodges-Kluck$^{1}$, Marc W. Pound$^{1}$, Andrew I. Harris$^{1}$, James
  W. Lamb$^{2}$ \& Mark Hodges$^{2}$}
\affil{$^{1}$Department of Astronomy, University of Maryland, College Park,
  MD 20742-2421\\$^{2}$Owens Valley Radio Observatory, California
  Institute of Technology, Big Pine, CA
  93513
}

\begin{abstract}
We report on Combined Array for Research in Millimeter-Wave Astronomy
(CARMA) and James Clerk Maxwell Telescope (JCMT) observations toward
the {\em Einstein} source 1E 1740.7-2942, a low-mass X-ray binary
(LMXB) commonly known as the ``Great Annihilator.'' The Great
Annihilator is known to be near a small, bright molecular cloud in a
region largely devoid of emission in $^{12}$CO surveys of the Galactic
center.  This region is of interest because it is interior to the dust
lanes which may be the shock zones where atomic gas from the HI
nuclear disk is converted into molecular gas.  We find that the region
is populated with a large number of dense ($n \sim 10^5$ cm$^{-3}$)
regions of excited gas with small filling factors.  The gas appears to
have turbulent support and may be the result of sprays of material
from collisions in the shock zone.  We estimate that $\sim 1 - 3
\times 10^5 M_{\odot}$ of shocked gas resides in our $r \sim
3^{\prime}$, $\Delta v_{\text{LSR}} = 100$ km s$^{-1}$ field.  If
this gas has recently shocked and is interior to the inner Lindblad
resonance of the dominant bar, it is in transit to the $x_2$ disk,
suggesting that a significant amount of mass may be transported to
the disk by a low filling factor population of molecular clouds with
low surface brightness in larger surveys.
\end{abstract}


\section{Introduction}

In spite of the difficulties of determining the structure and dynamics
of the inner 400 pc of the Galaxy, remarkable progress has been made
by comparing surveys in different tracers of molecular gas.  These
surveys have revealed the average properties of the gas
\citep{bania77,liszt78,bal87,stark88,bitran97,oka98,tsuboi99,mart04}.
The gas is excited, dense, and exists in conditions unique in the
Galaxy, including high external pressures and strong magnetic fields.
The most prominent line emission comes from the various rotational
energy levels of $^{12}$CO, and surveys exist from $J=1-0$ to $J=7-6$.
It is evident from these degree-scale results that the molecular
medium is not well described by its average properties, but over the
past few decades a convincing picture of the Galactic center has
emerged.  At least one stellar bar crosses the Galaxy, and the $x_1$
and $x_2$ orbit families resulting from its potential contain atomic
and molecular gas respectively \citep{bin91}.  The molecular gas on
the $x_2$ orbits is largely found in a ``Galactic center ring'' (GCR)
and is both dense and turbulent \citep{fux99}.  Because gas on the GCR
is forming stars, it is unlikely that the molecular medium inside the
$x_1$ disk is primordial, and therefore some mechanism must exist for
converting atomic gas to molecular gas.  The region in between the
atomic disk and the GCR, the site of the bar's inner Lindblad
resonance (ILR), is where the transition occurs, and is thought to
contain two dust lanes in small spiral arms where molecule-forming
shocks and sprays occur \citep{rod06}.  The GCR and surrounding
molecular gas near these dust lanes are collectively known as the
``central molecular zone'' (CMZ).  The most obvious failure of this
model is that it has not yet explained the asymmetric distribution of
gas around the dynamical center of the Galaxy, Sgr A*, although
\citet{rod08} argue that the GCR may be accreting from only one 
side. 

This simple picture is remarkable considering the observed complexity
of the region.  However, it necessarily overlooks details critical to
a complete understanding.  In particular, we are interested in the
specific nature of the process converting atomic gas to molecular gas.
That the dust lanes are the site of molecule-forming shocks is likely
given the absence of star formation, and studies in line ratios
tracing shock chemistry indicate that the lanes contain shocked gas
\citep{rod06}, but the shock environment has yet to be explored in
detail.  Pre- and post-shock environments appear as relative voids in
$(l,v)$ diagrams; the emptiness of the region varies with the choice
of spectral line, but even in $^{12}$CO (1--0), it is clear that there
are empty regions within the CMZ, inside what \citet{bin91} saw as the
``CO parallelogram'' and earlier studies identified as an ``expanding
molecular ring.''  This boundary is likely several structures including
the molecule-forming shock zones \citep{rod08}, and since shocked gas must fall
farther into the Galactic potential, the relative paucity of emission
interior to the parallelogram is interesting.  There are several
possible explanations for the voids including (1) the shocked gas
primarily joins the $x_2$ disk at certain regions (e.g. the $l =
+1.3^{\circ}$ cloud), (2) the filling factor of shocked gas in these
regions is too low to be seen by large-scale surveys (along the lines
of the suggestion of \citet{oka98}), (3) the newly-formed molecular
bundles are extremely diffuse or not emissive, or a combination of the
above.  We expect that studying the post-shock side of the dust lanes
will provide information regarding the state in which the gas arrives
on the $x_2$ disk, and may also be a useful contrast to molecular
cloud formation in dust lane shocks in the disk.

\subsection{The Great Annihilator}

Motivated by published data \citep{bal91}, we chose to observe the area near a bright
molecular cloud near the far side dust lane at $l = -0.9^{\circ}$, 
$b=-0.1^{\circ}$, and $v_{\text{LSR}} = -140$ km s$^{-1}$ which lies in
such a void in $^{12}$CO (1--0) and HCN (1--0) surveys \citep{bal87,lee96}.
Historically, this cloud has been of interest because of its coincidence on
the sky with the low mass X-ray binary (LMXB) candidate known as the
``Great Annihilator'' ({\em Einstein} source 1E 1740.7-2942).  The Great
Annihilator is a bright X-ray and $\gamma$-ray source and was originally
thought to be a black hole accreting directly from a host molecular cloud
which, in turn, was brightened by the association \citep{bal91,mir91}.
The original hypothesis of Bondi-Hoyle accretion from the ISM is almost certainly
incorrect given the present understanding of the system as a LMXB
\citep{main99}, but as \citet{weid08} proposed a correlation in spatial
distribution of LMXB systems and diffuse $\gamma$-ray emission in the
Galaxy, the strong and narrow annihilation line still makes the possibility
of physical association intriguing.  Historically, the two strongest
pieces of evidence for such an association were the stated small 
probability ($\sim 5$\%) of a chance association with a bright molecular
cloud in the region \citep{bal91} and the apparent offset in HCO$^+$ 
(1--0) emission from the jets seen with the VLA 
\citep{phil95}.   This offset was attributed to the high levels of
ionization which would destroy the HCO$^+$ close to the black hole,
and \citet{lepp96} suggested that HCN may be able to survive in more
highly-ionized gas, proposing future observations.

Our interest in the gas dynamics of the inner Galaxy led us to consider
the Great Annihilator region as a target for a pathfinding observation
for two reasons: (1) the bright cloud is in a ``post-shock void'' region
and (2) it has a well documented brightness and spectrum in several
molecular lines \citep{bal91,mir91,phil95}.  This cloud is thus a reliable
pathfinder target for a larger map in the region near the shock zones.  It
is desirable to have a known bright target included in a wide field both
for orientation and determining the suitability of a given instrument to 
our science goals.  Additionally, even supposing the LMXB is responsible
for the {\em brightness} of the $v_{\text{LSR}} = -140$ km s$^{-1}$ cloud, it is
not responsible for the {\em presence} of the dense gas.  The detection 
of one bright cloud, therefore, suggests the presence of additional
molecular gas.  

\subsection{Typical Voids}

However, it is imperative to consider whether the small region near the Great
Annihilator is typical of the void regions we wish to study; determining
whether a relative void is ``typical'' requires choosing a scale.  We do this
by considering both an observational strategy and our broader science goals.
We have reason to expect the detection of emission in these voids based on the
publically-available \citet{bal87} $^{12}$CO (1--0) data.  In the published
$l-v$ maps of this data, which have been integrated over several 1$^{\prime}$
grid points in latitude, the relative voids appear empty, but an examination
of the $l-v$ maps for each slice in latitude shows faint molecular emission
in parts of these regions.  An absolute void across all latitude slices is
probably actually empty, but when integrating across several slices, a faint
structure with little extent in latitude will become fainter relative to
extended structures nearby.  For example, in the \citet{bal87} data between
$ -0.11^{\circ} < b < -0.076^{\circ}$, there appears to be a bridge of emission
between the CO parallelogram shock lanes and the $x_2$ orbits at $l=-0.3^{\circ}$,
$v_{\text{LSR}} = 120$ km s$^{-1}$.  This emission occurs in a region we would
expect to be mostly empty according to bar parameters such as those in 
\citet{rod08}, but may be related to the highly excited bundles of gas seen 
in earlier \citet{jen94} work.  Closer to the Galactic plane, this bridge does
not appear, whereas the surrounding $x_2$ disk and shock lane structures 
persist at the same longitudes; the region appears in the integrated map
only as a void.  Interior to the CO parallelogram, a ``typical'' region for
detailed study ought to be large enough to probe both the apparently empty
regions as well as some of the faint emission.  

For a detailed interferometric study of the region, the scale should clearly
be much larger than an individual molecular cloud's size, and we postulate
that the projected radius of the cloud associated with the Great Annihilator
is typical.  Assuming interaction with the LMXB as well as some intrinsic
brightness, this {\em measured} radius is only influenced by the presence of
the LMXB to the extent that the LMXB is responsible for the brightness of
the cloud, i.e. the ratio between excitation produced by the Great Annihilator's
particle emission and that produced by other means.  Based on the evidence
for a well-defined spectral peak at $v_{\text{LSR}} = -140$ km s$^{-1}$
centered spatially at some distance from the LMXB, as well as the physical
arguments for how far the positrons could travel \citep{phil95}, it seems
likely that the measured radius is close to the physical one.  As discussed
in \S1.1, the LMXB may not be physically associated with the cloud at all.
The field of view must be large enough to isolate individual molecular
clouds of this size (or small structures), but small enough that good
$u-v$ coverage and sensitivity can be achieved over the entire field of
view. A happy spatial medium exists for fields several arcminutes in radius;
the known cloud has a radius on the order of 1 pc ($24.3^{\prime\prime}$ for
a distance of 8.5 kpc) and a field several arcminutes ($1^{\prime} \approx
2.5$ pc) on a side could accommodate many such clouds.  For the velocity
coverage, while constrained by the correlator for a given observation, we
must appeal to the $l-v$ maps of the \citet{bal87} data and the theoretical
models we wish to test \citep{bin91}.  An inspection of
the $^{12}$CO (1--0) data suggests that the easiest way to view a sizable
segment of the post-shock region at once is to observe near one of the
vertical segments of the shock lanes as projected in the $l-v$ plane thanks
to our large range in $\Delta v$ compared to $\Delta l$ or $\Delta b$.  We
also wish to avoid contamination from line-of-sight molecular emission and
larger scale structures of the inner Galaxy.  For this study, we mapped a region near the
Great Annihilator described by $r \sim 5^{\prime}$ and $\Delta v_{\text{LSR}}
\sim 100$ km s$^{-1}$ which meets these criteria. 

In \S2 we discuss in detail our
observations of the cloud and its environs with both the Combined
Array for Research in Millimeter-Wave Astronomy (CARMA) and the James
Clerk Maxwell Telescope (JCMT), in \S3 and 4 our results and analysis,
and in \S5 the implications of our results, including an assessment of
whether our region is ``typical.''

\section{Observations \& Data Reduction}

\subsection{CARMA D \& E Arrays}

The Combined Array for Research in Millimeter-wave Astronomy (CARMA)
is a heterogenous interferometric array made up of six 10.4 and nine
6.1 m radio dishes located at 2195 m at Cedar Flat in the Inyo
Mountains of California.  The observatory operates in several
different configurations.  We report here on two tracks in the compact
August 2007 E-Array and one track in the April 2008 D-Array (Table
\ref{obstable}) towards the Great Annihilator cloud in HCN $J=1-0$
(88.63 GHz) and HCO$^+$ $J=1-0$ (89.19 GHz).  For these tracks, we
made mosaicked maps towards, but slightly offset from, the peak of
emission in the \citet{phil95} HCO$^+$ (1-0) map at 17:40:43.0
-29:43:25.0 (1950.0).  The HCN and HCO$^+$ observations were conducted
simultaneously with both lines in the upper sideband and with a
velocity range of -90 to -190 km s$^{-1}$. For one of the E-array
tracks (baselines 8--66 m) and the D-Array track (baselines 8--108 m),
the weather was excellent, whereas the second
E-Array track produced usable data requiring substantial flagging.
Visibilities in the E-array which had one antenna shadowing another 
were discarded.  While the shortest baselines tracing the largest 
structures are lost in this procedure, the E-Array data considered by
itself nonetheless reproduces the D-Array detections wheree the fields
overlap.  Importantly, the E-array includes numerous shorter, 
non-shadowed baselines significantly different from those in the D-Array,
producing significantly improved $u-v$ coverage. 
To calibrate and analyze the data, we employed the MIRIAD
software package \citep{sault95}.  We customized a standard script
provided by CARMA to inspect the data, apply calibrations and flags,
and extract clean maps.  We used an SDI CLEAN algorithm to generate
the maps presented here; a check against a maximum entropy
deconvolution scheme showed no significant differences.

\begin{figure}
\begin{center}
\includegraphics[scale=0.85,angle=270]{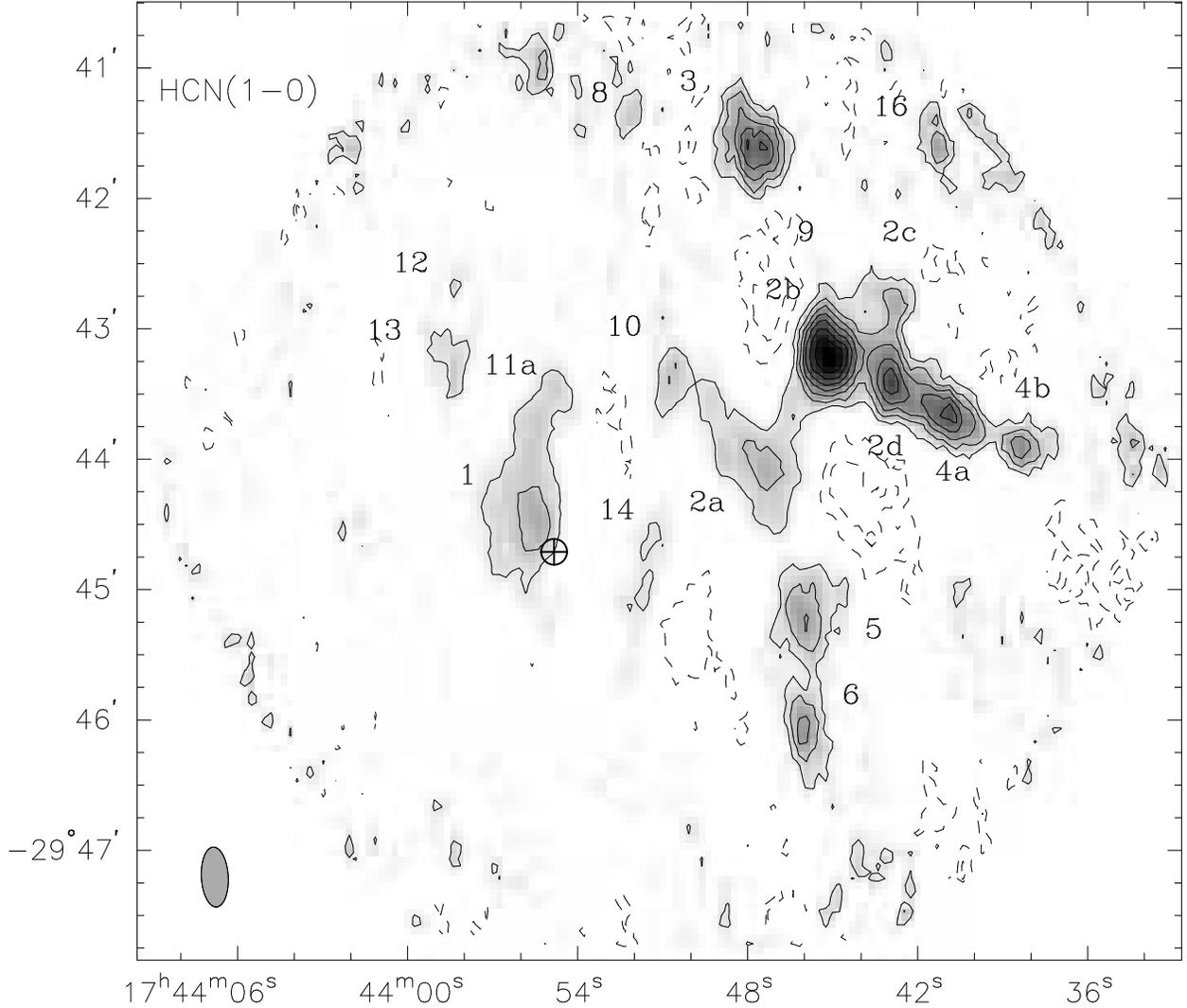}
\caption{Contour map of integrated amplitude of HCN (1--0) for combined
  D and E-Array data.  Dashed contours indicate negatives in the map.
  The mean RMS for the integrated map is 2.6 Jy/beam $\cdot$ km
  s$^{-1}$; a 1-contour feature is roughly a 2$\sigma$ detection.
  Some real detections appear weaker in the map as a result of spatial
  coincidence with negative features, others are behind brighter
  clouds--see text.  Clumps in Table \ref{measuredtable} have been labeled.
  A crosshair near cloud 1 shows the location of 1E 1740.7-2942
  \citep{mir91}, and the beamsize is at bottom left.  The
  Galactic equator is not in our field, but is to the top right.}
\label{hcnmap}
\end{center}
\end{figure}

\begin{figure}
\begin{center}
\includegraphics[scale=0.85,angle=270]{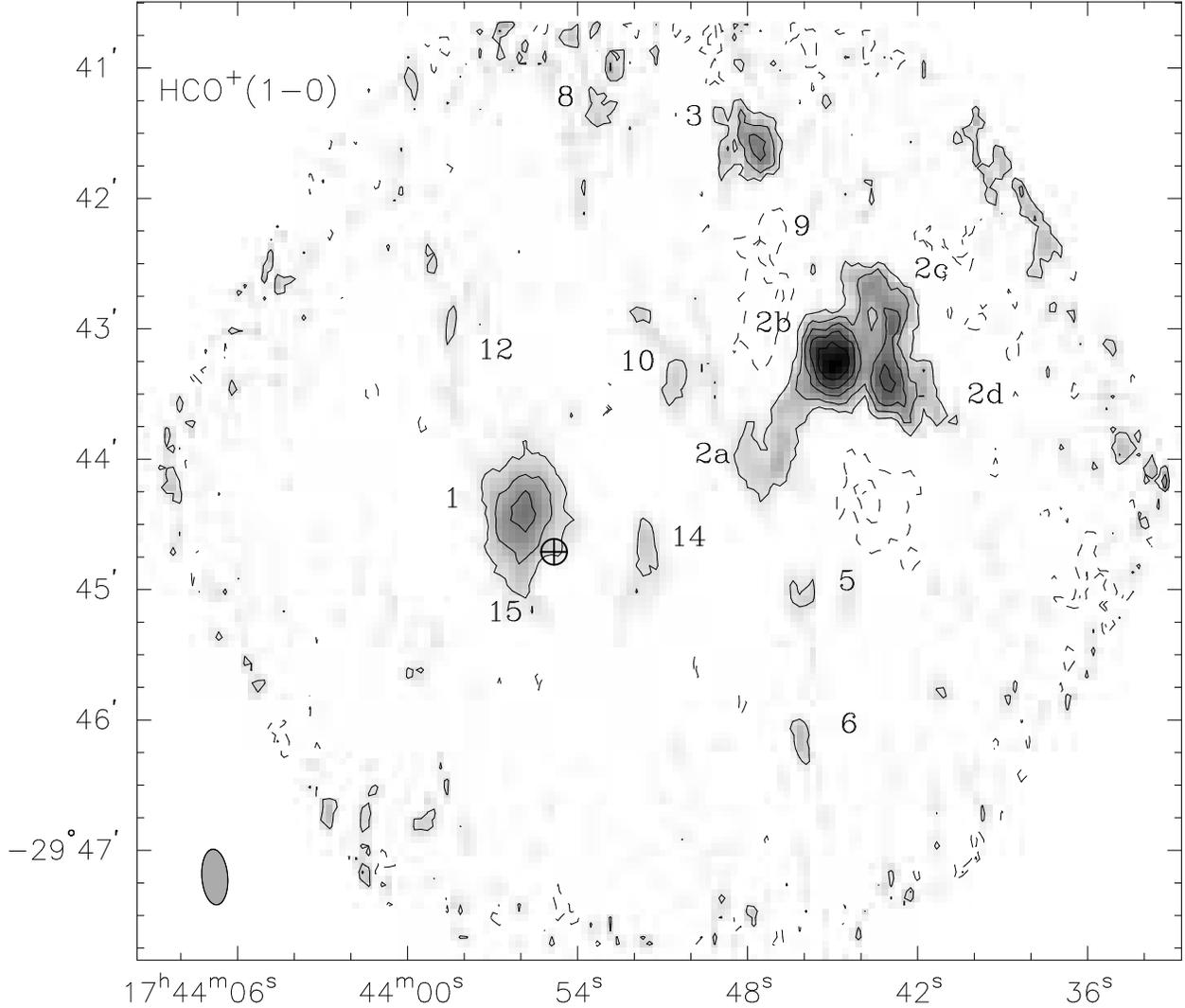}
\caption{Contour map of integrated amplitude of HCO$^+$ (1--0) for combined
  D and E-Array data.  Dashed contours indicate negatives in the map.
  The mean RMS for the integrated map is 1.7 Jy/beam $\cdot$ km
  s$^{-1}$; a 1-contour feature is roughly a 2$\sigma$ detection.
  Some real detections appear weaker in the map as a result of spatial
  coincidence with negative features, others are behind brighter
  clouds--see text.  Clumps in Table \ref{measuredtable} have been labeled.
  A crosshair near cloud 1 shows the location of 1E 1740.7-2942
  \citep{mir91}, and the beamsize is at bottom left.  The
  Galactic equator is not in our field, but is to the top right.}
\label{hcomap}
\end{center}
\end{figure}

Figs. \ref{hcnmap} \& \ref{hcomap} show the integrated amplitude maps
for the combined calibrated D \& E-Array data, and it is immediately
obvious that there is a large number of features, some of which are
spatially resolved, and that many of these features are weak or absent
in the HCO$^+$ data.  The clouds are dense and the spatial coverage of
our $r \sim 3^{\prime}$ field between -90 and -190 km s$^{-1}$ is
about 25\%.  At a glance, it is clear that we see a large number of
features which may have small filling factors in larger beams or
survey grids.  We clearly detect the Great Annihilator cloud (labeled
cloud 1) to the left of map center.  Note that because there were
fewer E-Array mosaic pointings than in D-Array (Table \ref{obstable}),
the SNR decreases towards the map edge.  Because the map contains
negative amplitudes, and features at different velocities may overlap
spatially, the integrated maps do not accurately reflect the strength
of emission in any one cloud; we found clumps that have no labels in
Figs. \ref{hcnmap} \& \ref{hcomap}. These clumps are evident in the
channel maps (Figs. \ref{hcnchannel} \& \ref{hcochannel}).

\subsection{James Clerk Maxwell Telescope}

\begin{figure}
\begin{center}
\includegraphics[scale=0.80]{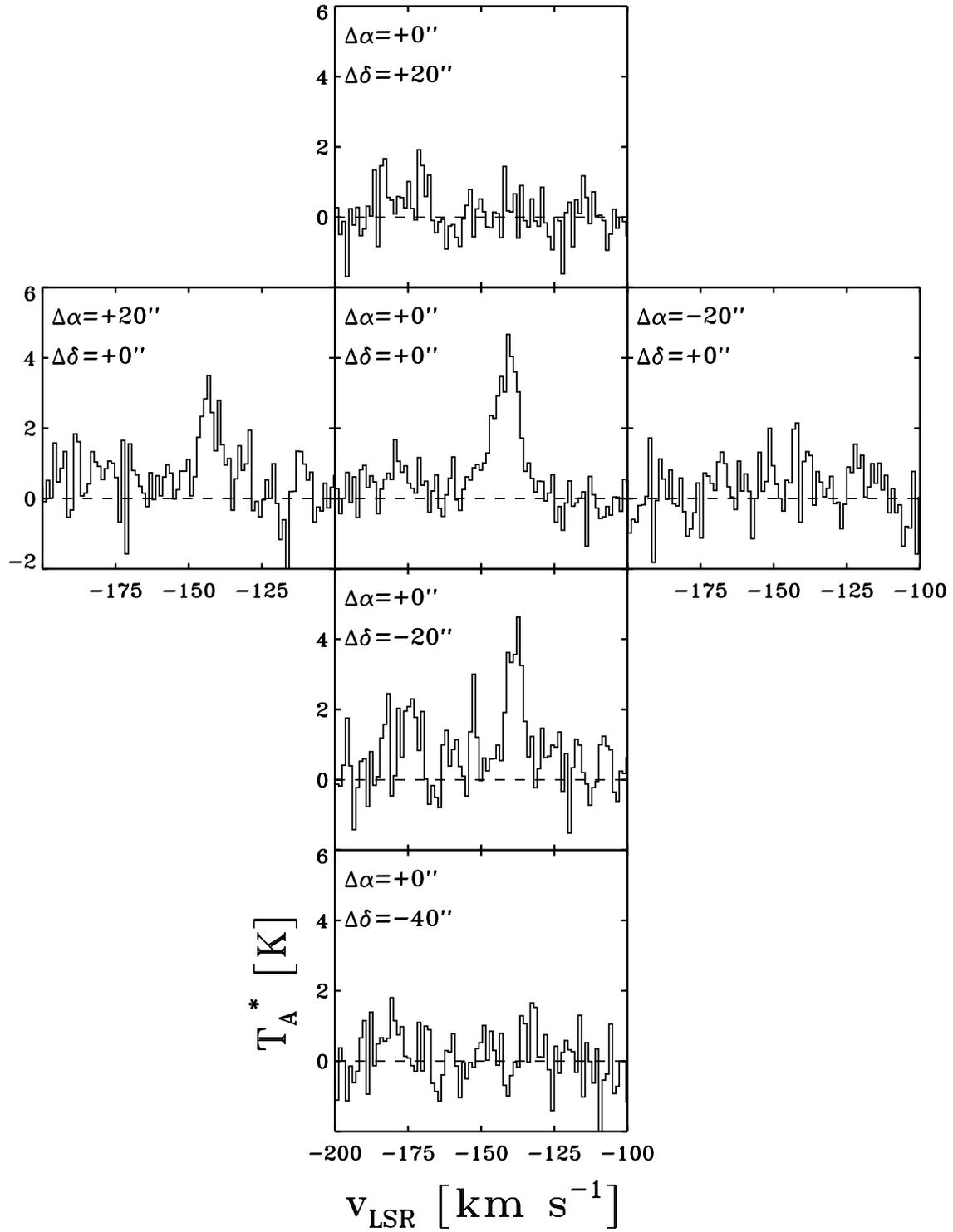}
\caption{JCMT $^{12}$CO $J=6-5$ map towards peak of \citet{phil95} on
  cloud 1 near the Great Annihilator.}
\label{jcmtmap}
\end{center}
\end{figure}

We obtained $^{12}$CO $J = 6-5$ data for the map in Fig. \ref{jcmtmap}
on 1995 April 23 with the 15 m James Clark Maxwell Telescope located
at the Mauna Kea Observatory.  We used the FANATIC submillimeter
spectrometer \citep{harris94}, a quasioptial receiver with a
solid-state LO and an SIS mixer.  For this observing run the mixer
device and associated dual-slot antenna was supplied by J.~Zmuidzinas;
the receiver temperature was about 400~K (DSB).  The weather was
excellent, with 55\% zenith transmission at the line's 691~GHz
frequency, deduced from measurements of the sky's emission
temperature.  The intensity scale for Fig. \ref{jcmtmap} is $T_A^*$
for a forward coupling to a Jupiter-size ($\sim 40^{\prime\prime}$)
source.  Further information on observing and calibration methods may
be found in \citet{harris95}.

The central field of Fig. \ref{jcmtmap} is pointed towards the peak of
HCO$^+$ (1--0) emission in \citet{phil95}, and the remainder of the
spectra are taken from fields offset by 20$^{\prime\prime}$ in
$\alpha$ and $\delta$; each pointing lasted for 240 s.  Additional
data were obtained in $^{13}$CO(6--5) toward the Phillips \& Lazio
peak, but no line was detected with an upper limit 20 times smaller
than the $^{12}$CO (6-5) line brightness toward the same position.

The detection of the cloud in the $J=6-5$ line demonstrates that the
cloud is both warm and dense.  The approximate size and shape of the
cloud traced by Fig. \ref{jcmtmap} is in agreement with the CARMA maps
and the \citet{mir91} HCO$^+$ (1--0) map.

\section{Observed Properties}

The presence of a number of bright regions in Figs. \ref{hcnmap} \&
\ref{hcomap} and the suggestion that these regions may be quite
excited is an important result.  To measure  the properties of these
regions, we used the {\tt clumpfind} algorithm \citep{williams94}
using the average rms noise per spectral channel within $3^{\prime}$
of map center and requiring 60 pixels per clump (about twice the
area).  Clumps found at  $r > 3^{\prime}$ reside in the low-signal map
edge, and were therefore rejected, although we found additional
structure in the map edge we do not further discuss here.  The {\tt clumpfind}
results are given in Table \ref{measuredtable} and are consistent with
a visual inspection in that the same regions are identified.  Because
{\tt clumpfind} treats pixels along the $v_{\text{LSR}}$-axis no
differently than the spatial pixels, many of the clumps reported are
spatially unresolved at the $2\sigma$ level, and in extracting spectra
for these cases we use the beamsize instead of the reported {\tt
clumpfind} radii.  In Fig. \ref{spectra} we show the spectra extracted
for these clumps; for convenience, where {\tt clumpfind} found
multiple clumps which appear to be associated along all three axes we
have extracted one average spectrum for the region using the MIRIAD
{\tt imspect} task.  These spectra were extracted from roughly the
2$\sigma$ contour tracing the association boundaries.  Using {\tt
clumpfind} introduced a systematic bias into the sizes of the clumps
along the spatial and $v_{\text{LSR}}$-axis, and we discuss the
implications of this bias on our derived masses in \S4.1.

\subsection{Average Description}

For our analysis, we divide our clumps into resolved associations and
unresolved individual clumps, where the former are identified as
groups of clumps in the same region of $\alpha$, $\delta$, and
$v_{\text{LSR}}$.  The resolved clumps have characteristic scales of
$r \sim 1$ pc ($24.3^{\prime\prime}$ at a distance of 8.5 kpc) and, in
some cases, several resolved regions may themselves be associated; we
consider cloud 2 to be a molecular cloud containing multiple dense
regions since we see  nearby negative dips at similar peak velocities.
Hereafter we adopt ``cloud'' to refer to resolved associations of
clumps found with {\tt clumpfind} and ``clump'' to refer to any
individual component or unresolved feature.  The resolved clouds are
labeled 1, 2, 3, 4, 5, and 6, and these are the brightest features in
the map; the fainter clouds are unresolved, but may have extended
structure we would see with a deeper integration.  In the case of
clump 2b, the algorithm found two clumps in the HCN data which are
spatially coincident and differ by 5 km s$^{-1}$ along the
$v_{\text{LSR}}$-axis, but a detailed inspection of the channel maps
(Figs. \ref{hcnchannel} \& \ref{hcochannel}) convince us that {\tt
clumpfind} ought not to have found two clumps.  Similarly, two clumps
were found in cloud 3 where a visual inspection suggests that they
should be considered one structure.  The clumps are listed as the
algorithm found them in Table \ref{measuredtable}, and we use the
results from the algorithm through the rest of the analysis, but we
note that there may be systematic inaccuracies with our approach.  All
other clumps found by {\tt clumpfind} passed visual inspection.

The HCN and HCO$^+$ line strength is comparable in about half of the
clumps; perhaps interestingly, only cloud 1 has a significantly
stronger HCO$^+$ line.  On average, where HCO$^+$ emission is present,
the FWHM line widths are comparable to the HCN despite differing line
strengths.  The lines are roughly symmetric, but we lack the signal to
rigorously investigate the line profiles.  Measured FWHM line widths
for individual clumps range from 3--14 km s$^{-1}$ whereas average
line widths for the clouds as presented in Fig. \ref{spectra} range
from 8--30 km s$^{-1}$, assuming the line is adequately represented by
a single Gaussian component.

As expected, many of the clumps and associations are seen within the
faint emission described in \S1.2 (Fig. \ref{lvmap}).  
Clump 1, associated with 1E
1740.7-2942, is closer to the $x_2$ orbits and may not be part of 
the same structure.  Another group of clumps lies near -180 km s$^{-1}$
and may lie on the $x_2$ orbits, but the grouping in $l$ and $b$ 
suggests that these clumps are actually on non-circular orbits, since
they are grouped between $-0.10^{\circ} < b < -0.08^{\circ}$ and there
is a relative {\em lack} of emission where the GCR intersects our
region in the \citet{bal87} data.  The apparent clustering in Fig. \ref{lvmap}
is largely a result of plotting all the clumps found with {\tt clumpfind}
and, when associations such as cloud 2 are considered as one object
at some average $(l,v)$, the clumps appear randomly distributed (in 
latitude, they are distributed preferentially towards the Galactic equator,
and we do not have data for $b > -0.03^{\circ}$).  If the clouds were on $x_2$ 
orbits, we would expect them to be clustered towards larger negative
$v_{\text{LSR}}$ near -180 km s$^{-1}$ at $l \approx 0.9^{\circ}$. 
The isolated clumps and associations of gas are isolated from one
another along all three axes such that it seems unlikely that they
are part of one larger complex.  This implies that the ``bridges''
seen in $l-v$ slices of the \citet{bal87} data are not coherent
structures, but rather, the coherent structures exist on smaller
scales.  We remark on the individual clouds below.

\subsection{Resolved Clouds}

\paragraph{1: -140 km s$^{-1}$} Cloud 1 is the molecular cloud associated with
1E 1740.7-2942 and is the only cloud in the sample for which the
HCO$^+$ line is much stronger than the HCN line.  In the CARMA maps,
we do not see the extended structure to the south which \citet{phil95}
attribute to different ionization regimes near the jets of 1E
1740.7-2942, but we do see some low-signal emission to the southeast
(clump 15) at about -130 km s$^{-1}$ which may be the ``ridge'' they
describe.  The JCMT map does not extend far enough to assess the
presence of clump 15 in CO $J=6-5$, but clearly shows that the
highly-excited CO traces the HCO$^+$ well in the region observed and
that the cloud is compact.  Clumps 11a and 11b at -180 km s$^{-1}$ are
spatially coincident with cloud 1, and may be physically associated
with each other.  In \S5.2 we discuss how these results bear on a
possible physical association with 1E 1740.7-2942.

\paragraph{2: -100 km s$^{-1}$} The clumps labeled 2a-2d are each resolved
individually, but are close on the sky, have nearly identical peak
velocities, and are near strong negatives in the deconvolved maps
which also appear at the same velocity.  These negatives indicate
large-scale structure resolved out by the interferometer and motivate
a physical picture of several dense clumps within a diffuse envelope.
The HCN and HCO$^+$ line strengths are roughly equal in each cloud,
and 2b has the brightest peak flux in the map (Fig. \ref{spectra}
shows the HCO$^+$ in cloud 1 as stronger because it is summed over an
extraction box).  However, cloud 2b coincides on the sky with a
smaller clump at  -180 km s$^{-1}$ which makes the integrated
amplitude in the region in Figs. \ref{hcnmap} and \ref{hcomap}
significantly brighter than the contribution from 2b alone.  Since the
HCO$^+$ emission is weaker than the HCN, 2c is not identified as a
clump in the HCO$^+$ data and is rather split into two tails extending
from 2b and 2d.  We assume that because it is identified as a separate
clump in HCN that it ought to be identified separately from the two
brighter emission cores.

\paragraph{3: -115 km s$^{-1}$} This cloud is curious, since the average
FWHM line width over the association is significantly smaller in
HCO$^+$ than HCN despite the strong emission in both lines.  This
effect is noticeable in the individual clumps found in the region, and
is pronounced even when the spectrum is extracted from a box within
contours $>4\sigma$, indicating that perhaps the HCO$^+$ in the cloud
is confined to the densest regions.

\paragraph{4 \& 6: -125 km s$^{-1}$} Clouds 4a and 4b have similar
peak velocities and line widths in their average spectra, and both are
much weaker in the HCO$^+$ line than the HCN.  The clouds are notable
for their broad lines which {\tt clumpfind} breaks up into strings of
smaller clumps at similar velocities.  The similar peak velocities
mean that 4a and 4b are physically separated by $<1^{\prime}$ and
suggest they are or were part of the same complex, although we do not
see negative amplitudes that would point to a diffuse envelope.  Cloud
6 is similar in line width, size, and line strength to either
component of cloud 4, but it is several arcminutes away.

\paragraph{5: -180 km s$^{-1}$} This cloud is close on the sky to cloud 6 and
is similarly weak in HCO$^+$, so the extracted spectrum includes a
small portion of cloud 6 which appears as a weak feature at -125 km
s$^{-1}$.  Cloud 5 is the only certainly resolved cloud at -180 km
s$^{-1}$, although clump 17 behind cloud 2b may barely be resolved.

\subsection{Unresolved Clumps}

As mentioned above, several clumps listed in Table \ref{measuredtable}
appear in the channel maps and not in the integrated amplitude maps.
This occurs either when the clump coincides with another cloud on the
sky (as in the case of clump 17 behind 2b) or with a negative at a
different $v_{\text{LSR}}$ (as in the case of clump 9).  The
``unresolved'' clumps do not appear to be a distinct population from
the resolved clouds, and are instead characterized by weaker emission.
We extracted average spectra from  regions of about a beamsize; {\tt
clumpfind} does find smaller spatial sizes for these clumps than for
the ones comprising resolved associations.  The line widths of the
unresolved clumps are in agreement with those of the resolved ones.

\section{Derived Properties}

We have detected $\sim 20$ clumps of gas in a 3$^{\prime}$-radius
field.  These clumps have $r \leq 1$ pc (assuming a distance of 8.5
kpc) and may reside in clouds with radii of a few parsecs.  The clumps
have line widths, fluxes, positions, and peak velocities that make it
difficult to characterize them as different parts of the same
structure.  Integrating across $\Delta v = 100$ km s$^{-1}$, the
clumps cover about 25\% of our field.  The gas tends to emit more
strongly in HCN than HCO$^+$, and at least one of these clouds (cloud
1) produces a $^{12}$CO (6--5) line which traces the HCO$^+$ and HCN
gas quite well, meaning the cloud is highly excited. Surveys in
$^{12}$CO (4--3) emission show that clouds near our field are excited,
although we know from comparing the JCMT $^{12}$CO (6--5) and
unpublished $^{12}$CO (2--1) data that not all gas in the region is so
highly excited.

We now use these measured properties to derive additional quantities
for our clouds, and on the basis of these results, propose that the
sources are dense ($n_{\text{H}_2} > 10^5$ cm$^{-3}$), and that they
have shocked recently.  Because of the weakness of the HCO$^+$ in many
of the clumps, our arguments regarding the dense gas rely on the HCN
data.

\subsection{Mass \& Density Estimates}

Mass estimates invariably reflect their underlying physical
assumptions, so further testing of these assumptions is required to
assess their accuracy.  The virial mass in particular is very
uncertain for Galactic center clumps.  Although the external pressures
appear to be an order of magnitude too low to bind clumps on our
scales \citep{miyazaki00}, the clumps may be bound by the strong
magnetic fields known to exist in the Galactic center, or may be
unbound.  The virial mass is primarily useful as a way to compare our
results to other work in the absence of better estimators but adopting
values from previous work is fraught with peril, since the structure
observed is necessarily dependent on the beamsize.

\begin{figure}
\begin{center}
\includegraphics[scale=0.8,angle=0]{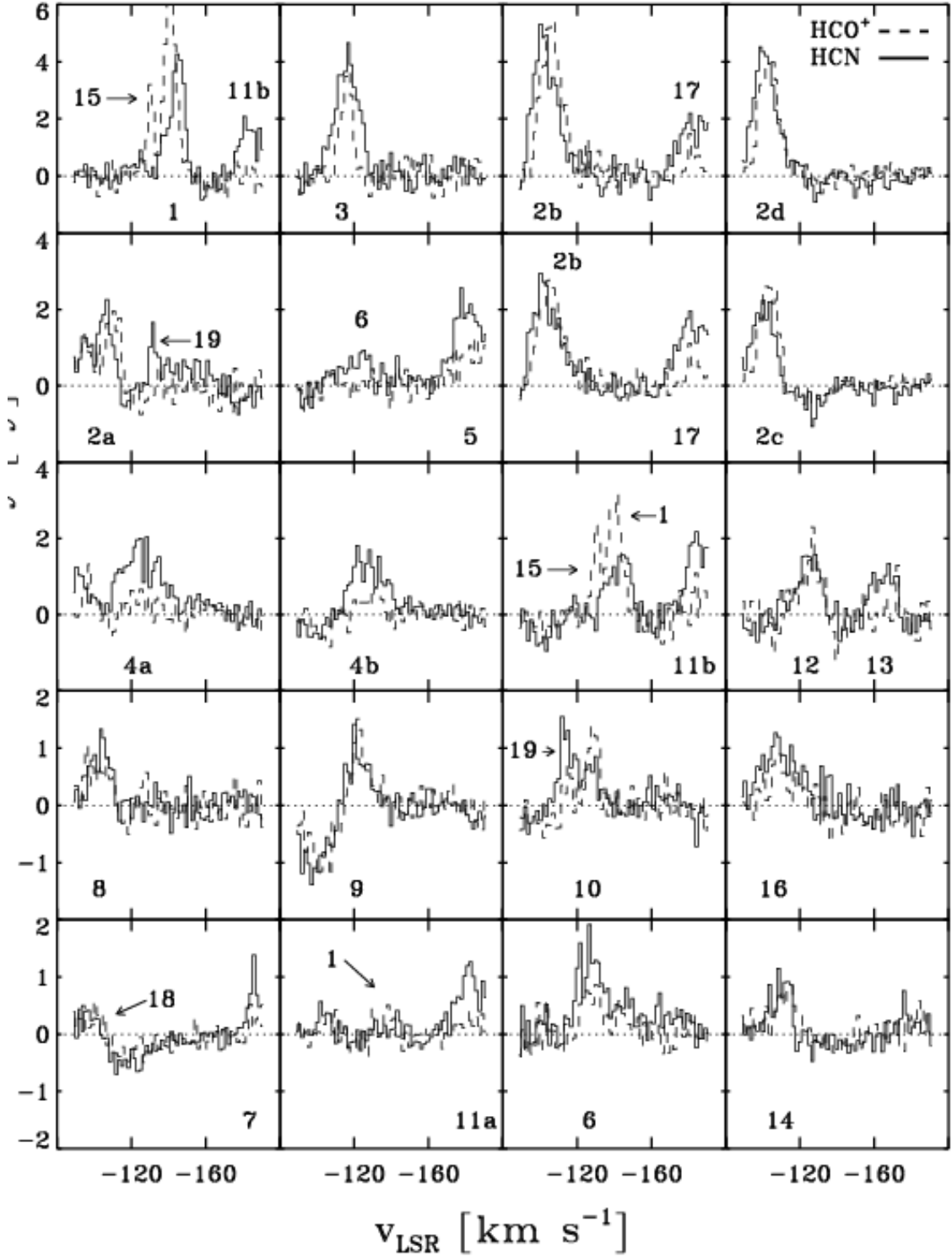}
\caption{Spectra extracted for clumps and clouds listed in Table
  \ref{measuredtable} from rectangular boxes roughly corresponding to the
  clump size at the 2$\sigma$ contour.  The spectra were extracted
  from the combined D and E-Array data in each line.  The label at the
  bottom of each frame marks the clump or cloud for which the spectrum
  was extracted; the same feature in a different frame has lower
  signal.  In cases where the 2$\sigma$ contour was smaller than a
  beamsize, the spectrum was extracted from a rectangle with the
  height and width of the major and minor axes of the synthesized beam.}
\label{spectra}
\end{center}
\end{figure}

\begin{figure}
\begin{center}
\includegraphics[scale=0.8,angle=270]{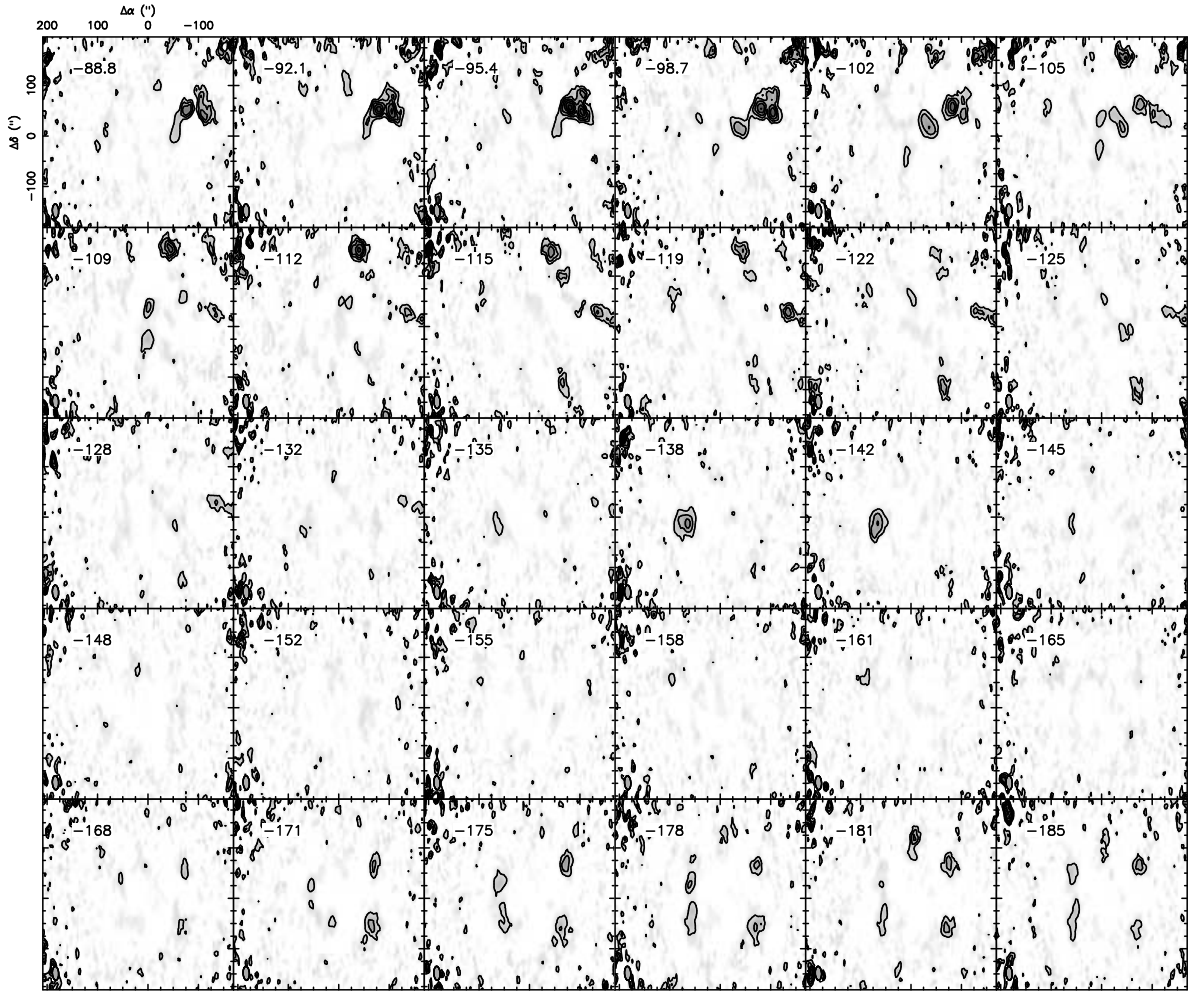}
\caption{Velocity channel maps for the combined D and E-Array HCN
  (1-0) data.  The 63 original channels have been binned into 30
  channels here.}
\label{hcnchannel}
\end{center}
\end{figure}

\begin{figure}
\begin{center}
\includegraphics[scale=0.8,angle=270]{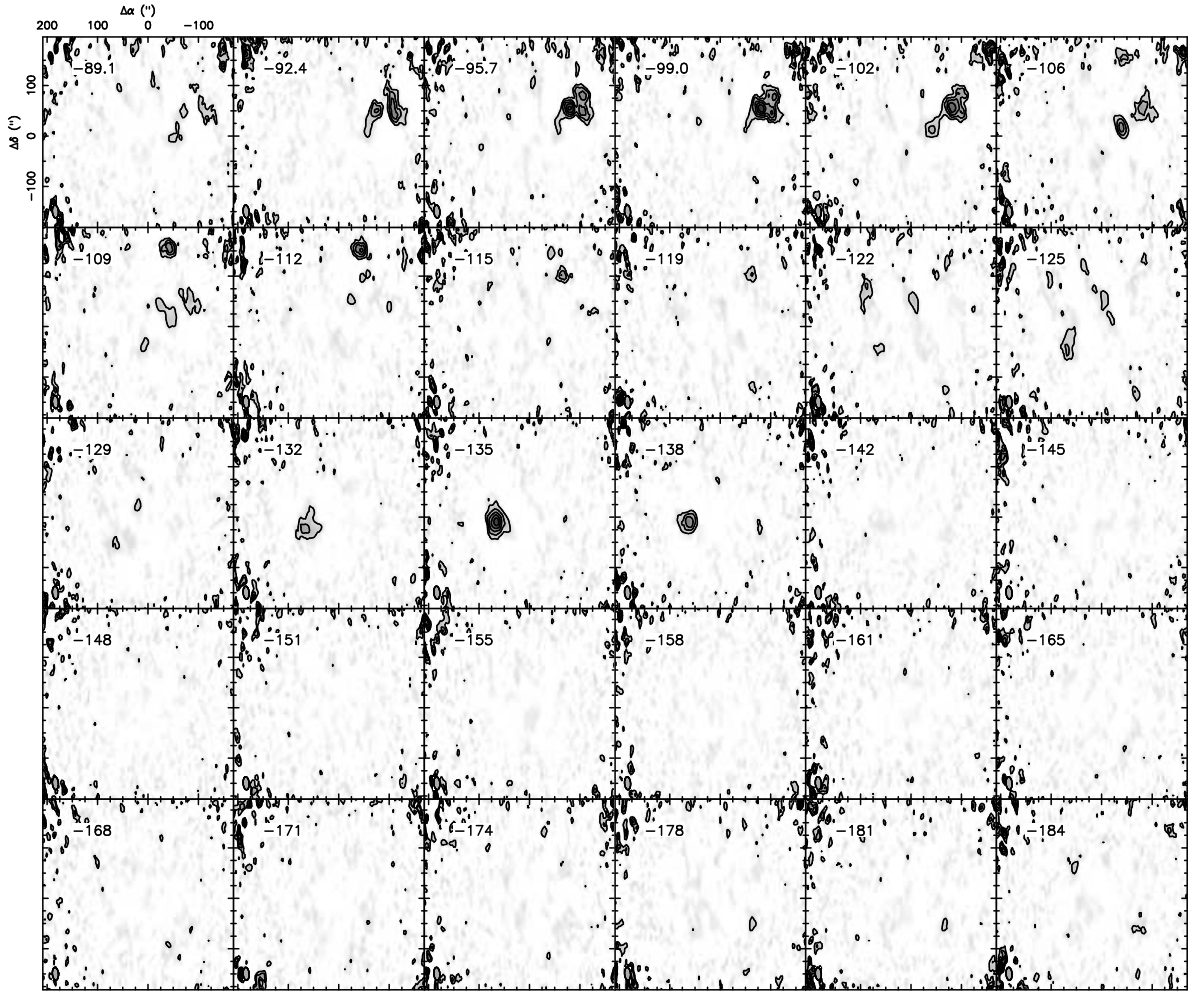}
\caption{Velocity channel maps for the combined D and E-Array HCO$^+$
  (1-0) data.  The 63 original channels have been binned into 30
  channels here.}
\label{hcochannel}
\end{center}
\end{figure}

\begin{figure}
\begin{center}
\includegraphics[scale=0.8,angle=0]{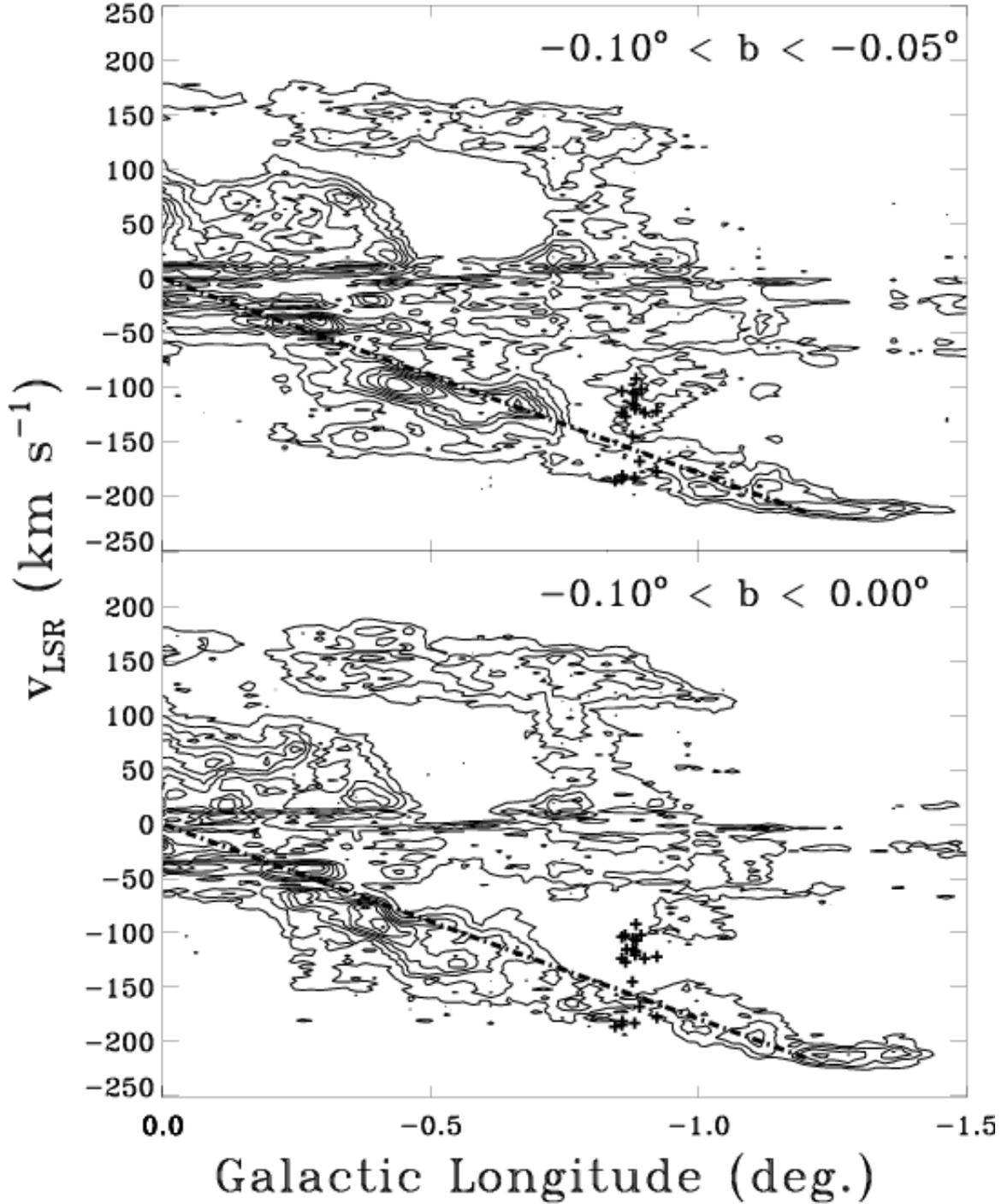}
\caption{\citet{bal87} $^{12}$CO (1--0) $l-v$ maps for (a) 
$-0.10^{\circ} < b < -0.05^{\circ}$ and (b) 
$-0.10^{\circ} < b < 0.00^{\circ}$ with crosses denoting the
coordinates of peak emission for our clumps 
as determined by {\tt clumpfind}.  The diagonal dashed line
indicates the approximate location of the $x_2$ orbits at 
this latitude range.  Contours
range from 2--30 K in intervals of $\Delta T = 2$ K.}
\label{lvmap}
\end{center}
\end{figure}

The sensitivity of the virial mass to line width means that whether
we take $M_{\text{vir}} = \Sigma_i m_{\text{vir},i}$, where a cloud is
made up of clumps with mass $m_{\text{vir,i}}$, or $M_{\text{vir}} =
\bar{\sigma}_v^2 R/G$, where $\bar{\sigma}_v$ is taken from an average
description such as Fig. \ref{spectra}, will influence our results.
The line width--size relation \citep{larson81,miyazaki00} cannot be
used to resolve such ambiguity because in our case it is systematic.
Reported line widths also generally appear to reflect the beamsize
used.  We investigated the extent to which noise influences {\tt
clumpfind} in order to determine whether multiple clumps  found in
resolved clouds accurately describe substructure.  There is almost
certainly substructure present: the question is whether {\tt
clumpfind} accurately detects it.  A visual inspection of the channel
maps in addition to testing {\tt clumpfind}'s behavior with artificial
resolved clouds in a featureless corner of our map convinced us that
the two clumps found in both 2b and 3 in the HCN are not real, and
that a single clump is a better description.  On the other hand, using
the line width from the average spectra in Fig. \ref{spectra} likely
overestimates the mass more severely than {\tt clumpfind}
underestimates it.  We therefore report the {\tt clumpfind} results
and posit that the masses and densities contained in the resolved
clouds are somewhat higher.  Table \ref{derivedtable} contains the
results for each individual {\tt clumpfind} detection,
$m_{\text{vir},i}$.  In determining $\bar{n}_{\text{H}_2}$ for each
clump, we used $\mu = 1.4 m_{\text{H}}$.

Is using the virial mass reasonable?  We know the clouds must be
dense, since typical critical densities for HCN (1-0) are $n \geq
10^5$ cm$^{-3}$, so the assumption that they are gravitationally bound
is not wholly unreasonable, although a shock compression might induce
similar densities which would then dissipate.  The line widths
indicate macroscopic turbulent support and are roughly symmetric, so
if the clouds are bound, they are unlikely to be much more
concentrated than at virialized relaxation.  We posit that for dense,
non-collapsing gas the virial mass is an adequate order of magnitude
description, but note that other mass estimates from studies with
larger beamsizes disagree with the virial mass by up to an order of
magnitude \citep{miyazaki00}.

In the resolved clouds our derived densities range from $10^{4.4-5.6}$
cm$^{-3}$ and, as we stated, we believe our results understimate the
{\em virial} mass.  If the LTE mass is a better descriptor of the
``real'' mass \citep{miyazaki00}, then our densities are $\sim 10$
times too high and, therefore, typical for Galactic center clumps with
radii 2--10 times larger than those of our clumps.  If our clumps are
smaller structures inside larger diffuse clouds such as those resolved
with larger beams, then they must either be much more dense than
$10^4$ cm$^{-3}$, or else comprise most of the mass and volume of
their hosts; for clumps $\sim 10$ times larger in {\em radius} the
latter seems unlikely.  The convergence of the critical density and
size arguments lends some credence to the virial estimates as an order
of magnitude estimate, so we believe our resolved clouds contain large
amounts of gas at densities exceeding $10^5$ cm$^{-3}$.  The
brightnesses and line widths of the unresolved clumps suggest that
they contain gas at similar densities, so we estimate that the total
mass contained in our cube is $1 - 3 \times 10^5 M_{\odot}$, contained
in parsec-scale bundles with a small filling factor.  We note that
since the HCO$^+$ line widths generally disagree with the HCN in a
given cloud, the HCO$^+$ virial masses would be different, but the
HCO$^+$ lines have a lower SNR, making the measurements more uncertain.

\subsection{Crossing Time}

Independently of the virial mass, we know the density to be quite high
from the critical density required to excite HCN (1--0).  In an
inspection of 2MASS J, H, and K-band images of our field, we do not see
enhanced star formation activity associated with our resolved clouds;
the clouds have not yet collapsed, so they are either unbound and will
decompress in a crossing time or bound and close to virialized.  In
either case, the clumps must be quite young if they are starless,
since they appear to have turbulent support.  The crossing time,
$t_{\text{cross}} \sim R/\Delta v_{\text{FWHM}}$, is about $10^5$ yr
(Table \ref{derivedtable}) for our resolved clouds.  Assuming a
circular orbit at 180 pc from the Galactic center with a velocity of
200 km s$^{-1}$ (the highest LSR velocities seen in the CO
parallelogram \citet{bin91}), the orbital period is $\sim 5 \times
10^6$ yr, over an order of magnitude greater than the dynamical
timescale. It is likely that the clumps have recently fallen onto
their present orbits, so they probably experienced a recent shock.
However, the constituents of cloud 2 suggest that condensation and
collapse may occur.

\section{Discussion}

\subsection{The Post-Shock Voids}

The \citet{bin91} hypothesis identifying the CO parallelogram with the
ILR of the Galactic bar may not be correct in detail \citep{fer07},
but the identification of the CO parallelogram with starless dust lane
shock zones \citep{rod06} strengthens the proposed mechanism for
molecule formation in shocks near the ILR.  The large surface density
of molecular gas in the Galactic center and orbit families in the
barred potential make it all but certain that continuous inflow from
the HI disk occurs. 

The density, excitation, and $(l,v)$ distribution of our clumps
suggests that we are indeed seeing molecular gas which has recently
formed in these shock zones.  Owing to their density, the clumps
represent a significant amount of mass ($>10^5 M_{\odot}$ in our field);
this mass is in a region where the orbits guarantee eventual transport 
to the $x_2$ disk.  This mass is transported in small bundles, possibly as
a spray of material from streamer collisions.  If these bundles are
unbound and decompress, diffuse molecular gas may rain down onto the
GCR, otherwise they may experience ballistic impacts with the diffuse
molecular envelopes in the GCR.  The small filling factor of the dense
regions may also explain the voids seen in larger surveys, although,
as noted in \S1.2, larger-scale diffuse emission was
detected in the \citet{bal87} maps in the region.  We expect that 
regions of similar size in the CMZ where faint emission is hidden by 
integration across latitude slices should contain gas similar to what
we see.  Since the gas must be on self-intersecting orbits, the voids
may also imply a characteristic timescale for clumps to fall to the
$x_2$ disk.

There are two challenges to the proposal outlined above.  First, the
clumps and clouds may be associated by a larger, bridge-like structure.
This would imply a directed mass flow at certain points on the CO
parallelogram rather than a large number of small bundles of shocked
gas distributed throughout the region in between the shock lanes and
the $x_2$ orbits.  Second, the multiple constituents of cloud 2 in a
larger envelope (but still small on the scales of large surveys) 
suggest that condensation into dense clumps, as seen in Galactic 
molecular clouds, may take place under higher pressures.  The first
case is difficult to assess because a close inspection of individual
latitude slices of the \citet{bal87} $^{12}$CO (1--0) data shows many
small features in between the CO parallelogram of \citet{bin91} and
the GCR.  We consider it unlikely that these represent steady-state
channels for mass flow because of the nature of the self-intersecting
$x_1$ orbits interior to the ILR.  However, it is possible that 
material from molecular and atomic gas clouds shredded in the shock
lanes maintains some coherent structure as it dissipates angular
momentum.  The second challenge may fit neatly into the first if larger
structures fall onto the self-intersecting $x_1$ orbits, but aside
from the constituents of cloud 2, we see little evidence for dense
bridges between otherwise distinct clumps.  Furthermore, we see no
evidence for star formation, so even internal turbulent shocks must
have been recent.  If a mechanism for recent shocks is
required, the proximity to the shock lanes provides a natural 
explanation.

To determine whether the Great Annihilator region appears
typical, we applied the scale of our map ($r \sim 5^{\prime},
\Delta v_{\text{LSR}} = 100$ km s$^{-1}$) to other areas along 
the shock lanes labeled in \citet{rod06} using the \citet{bal87} data cube. 
For the diagonal portions of the
CO parallelogram, we stretched our scale in longitude and shrunk it in
$\Delta v$.  Ignoring the regions near $v_{\text{LSR}} = 0$ km s$^{-1}$ 
and obvious large scale structures such as the GCR, we find that the
1E 1740.7-2942 region for our observation is typical of the post-shock
regions in the $^{12}$CO (1--0) map.  The emission associated with this 
region in the \citet{bal87}
data appears to come from a structure covering $\Delta b \sim 0.05^{\circ}$,
$\Delta l \sim 0.15^{\circ}$, and $\Delta v \sim 100$ km s$^{-1}$.  We
think it unlikely that the structure is bridged to the $x_2$ orbits based
on a detailed inspection of the region at different emission contour 
intervals.  We therefore have no reason to believe that the presence of
the Great Annihilator in the region detracts from a gas-dynamical analysis.
However, we note that this analysis assumes that the {\em observed}
similarities between regions in the \citet{bal87} $^{12}$CO (1--0) maps 
correspond to {\em physical} similarities in the environments of the
shock lanes.  Although the region is small, our results are largely
consistent with the \citet{bin91} theory of molecule-forming shocks at
the ILR even though, in detail, the scenario is more complex.  If we are
seeing clumps that have recently shocked, it is harder to fit them into
alternative pictures explaining the CO parallelogram, e.g. an ``expanding
molecular ring'' or the \citet{stark04} proposal of a stalled ring of
gas accumulating from the true ILR farther away from the Galactic
center. 

We know that the GMCs on the $x_2$ disk are forming stars
from cores in clumps, but we do not know whether the clumps
primarily form via condensation or external perturbation, nor the
filling factor of small clumps in gas accreting onto these orbits.
\citet{jen94} find in their sticky-particle hydrodynamical models
that the steady-state $x_2$ disk contains a large number of strongly
shocked clumps of gas, although the decompression timescale suggests
that if the clouds are still dense, they must be at least gravitationally
bound if they experience no additional shocks interior to the dust lanes.
Their simulations also find that although there are places where shocked
gas preferentially joins the $x_2$ disk (e.g. the $l = 1.3^{\circ}$ near-
side molecular complex), bundles of gas fall in from the shock zone at 
the ILR at many angles.  \citet{jen94} admit that their results are only
marginally successful at reproducing the features of the Galactic center,
and subsequent models (e.g. \citet{rod08}) have done much better.  Yet
the results we present here suggest that at least the accretion of
material onto the $x_2$ disk in the form of dense bundles is still a
possibility.

\subsection{1E 1740.7-2942}

Our data provide useful contrasts to two of the arguments made
previously for a physical association between 1E 1740.7-2942 and cloud
1, but we cannot rule out a possible association.  The argument in
favor of physical association due to a small chance of coincidental
superposition of a black hole with a bright molecular cloud depends in
detail on choosing a cutoff brightness.  The 5-7\% chance previously
reported \citep{bal91,mir91} relied on studying the density of peaks
in molecular line surveys near the Galactic center using some
brightness criterion.  Without establishing a similar criterion, we
cannot directly compute the probability of a chance association.
However, the presence of a large number of clumps in the region with
similar line widths and a wide range of peak fluxes suggests that a
highly excited clump is common enough not to require association with
a black hole.  At approximately the 2$\sigma$ contour, the chance of
coincidental superposition with {\em any} clump in our field is 25\%.
That the HCO$^+$ is noticeably stronger in cloud 1 is unusual, but
there are smaller clumps where HCO$^+$ is stronger than HCN.  With a
larger velocity range, the chance of coincidental superposition at
arbitrary velocity may well increase.

The \citet{phil95} argument in favor of physical proximity relies on
the interpretation of a ridge of HCO$^+$ emission parallel to the VLA
jets associated with 1E 1740.7-2942 as evidence for an ionization rate
gradient.  Their ridge is significant ($> 2\sigma$) and separated from
the VLA jets by $15^{\prime\prime}$ with a peak velocity close to -140
km s$^{-1}$, suggesting that it is part of the same structure as cloud
1, shown in Fig. 2 in \citet{phil95}.  The primary detection of cloud
1 in their OVRO map agrees well with both our CARMA and JCMT data in
spatial extent, but we do not detect the southern ridge apparent in
their map.  Instead, we see clump 15 at -130 km s$^{-1}$ to the
south{\em east} of cloud 1 -- it is closer to the VLA jets than
$15^{\prime\prime}$.  An inspection of clump 15 in $(l,v)$ space shows
a clear separation between it and cloud 1.  Moreover, this clump is
detected in the JCMT $^{12}$CO (2--1) data (not shown) whereas we do
not see it in $^{12}$CO (6--5).  \citet{mir91} also detect the clump
in $^{12}$CO (2--1) and not in CS (2--1).  There is undoubtedly
emission to the south of cloud 1, but it is not as excited or warm as
cloud 1, nor do we see evidence for a ``ridge'' linking it to cloud 1.
These results call into question whether an ionization rate gradient
exists.  Furthermore, \citet{lepp96} argue, based on their model, that
if the HCO$^+$ ridge is caused by an ionization rate gradient, we
would expect to see HCN molecules surviving closer to the black hole.
There is no evidence in our maps that HCN emission associated with
cloud 1 is closer to the {\em Chandra} X-ray source than the HCO$^+$.
Our results make it more difficult to make the case for physical
proximity between  the Great Annihilator and cloud 1.

\section{Conclusions}

We have used CARMA D \& E-Array observations in tandem with JCMT data
to study the region near 1E 1740.7-2942, a relative ``void'' in
$(l,v)$ diagrams of $^{12}$CO survey data.  The most important result
of our work is the discovery that even in the regions with low average
emissivity, there is a substantial amount of mass ($> 10^5 M_{\odot}$)
contained in small, dense, excited regions implying a recent shock.
These bundles have scales of $r \sim 1$ pc and appear randomly
distributed in $(l,v)$ space, so shocks or sprays in the nearby dust
lanes of the CO parallelogram naturally explain our observations.  The
small filling factor of the dense bundles accounts for why they are
not seen or only faintly present in large surveys.  We investigate in detail the relationship
between cloud 1 and the LMXB 1E 1740.7-2942 and argue that the
probability of coincidental superposition with excited gas is higher
than originally estimates, and that we see no evidence for a proposed
ionization rate gradient; we can explain why cloud 1 is excited in the
JCMT map without invoking a black hole.

\acknowledgments

This research was funded in part by the Astronomy Department of the
University of Maryland at College Park.  The authors thank M.
Leventhal for useful discussion of the diffuse 511 keV annihilation
radiation in the Galactic plane, and R. Genzel \& J. Zmuidzinas for
motivation and support of the JCMT observations.  We used 7-m Bell 
Labs data made publically available by J. Bally.

Support for CARMA construction was derived from the states of
California, Illinois, and Maryland, the Gordon and Betty Moore
Foundation, the Kenneth T. and Eileen L. Norris Foundation, the
Associates of the California Institute of Technology, and the National
Science Foundation. Ongoing CARMA development and  operations are
supported by the National Science Foundation  under a cooperative
agreement, and by the CARMA partner universities.  This material is
based on work supported by the National Science Foundation under grant
numbers AST-0540450 and AST-0540399.

\clearpage

\clearpage

\clearpage

\clearpage

\clearpage

\clearpage

\clearpage

\clearpage
\begin{deluxetable}{lcccccllll}
\tabletypesize{\scriptsize}
\rotate
\tablecaption{CARMA Observational Parameters} 
\tablewidth{0pt}
\tablehead{
  \colhead{Array Config.} 
   & \colhead{Source Int. Time} & \colhead{\# Mosaic Points} & \colhead{Gain Cal.} &
  \colhead{Passband Cal.} & \colhead{Flux Cal.} & \colhead{HCO$^+$
    Beam} & \colhead{HCO$^+$ PA} & \colhead{HCN Beam} &
  \colhead{HCN PA} \\
  & \colhead{(hr)} &  &  &  &  & \colhead{(arcsec)} & \colhead{(deg.)}
  & \colhead{(arcsec)} & \colhead{(deg.)} 
}

\startdata
D08A & 2.36 & 37 & 1733-130 & 2148+069 & MWC349 & 15.4 $\pm$ 0.5 $\times$
6.8 $\pm$ 0.2 & 1 $\pm$ 2 & 16.3 $\pm$ 0.6 $\times$ 7.0 $\pm$ 0.2 & 1
$\pm$ 2 \\
E07 & 4.43 & 7  & 1733-130 & 1751+096 & MWC349 & 27 $\pm$ 1 $\times$ 12.8
$\pm$ 0.5 & 5 $\pm$ 2 & 29 $\pm$ 2 $\times$ 13.4 $\pm$ 0.6 & 3 $\pm$ 2 
\enddata

\tablecomments{\label{obstable} Beamsizes and position angles determined
  using the MIRIAD {\tt mospsf} task within the central
  $20^{\prime\prime} \times 20^{\prime\prime}$ of the combined data
  sets.}
\end{deluxetable}

\clearpage

\begin{deluxetable}{llcccccc}
\tabletypesize{\scriptsize}
\tablecaption{CARMA Measured HCN/HCO$^+$ Clump Properties}
\tablewidth{0pt}
\tablehead{
  \colhead{Label} & \colhead{Line} & \colhead{$\alpha$} & \colhead{$\delta$
    } & \colhead{$v_{LSR}$} & \colhead{$T_{peak}$} &
  \colhead{$R_{\text{clump}}$} & \colhead{$\Delta v_{FWHM}$} \\
  & & \colhead{(J2000)} & \colhead{(J2000)} &
    \colhead{(km s$^{-1}$)} & \colhead{(K)} & \colhead{(arcsec)} &
    \colhead{(km s$^{-1}$)} 
}

\startdata

1   & HCO$^+$ & 17 43 56.1 & -29 44 19.0 & -141.7 & 2.47 & 26.9 & 4.8\\
1   & HCN     & 17 43 55.8 & -29 44 22.0 & -145.0 & 1.61 & 24.1 & 4.1\\
2a  & HCO$^+$ & 17 43 48.9 & -29 43 25.0 & -105.6 & 1.90 & 25.4 & 9.5\\
2a  & HCN     & 17 43 48.2 & -29 43 55.0 & -107.0 & 1.43 & 25.0 & 10.6\\
2b\tablenotemark{1}  & HCO$^+$ & 17 43 44.8 & -29 43 16.0 & -103.9 & 2.51 & 26.9 & 4.8\\
2b  & HCN     & 17 43 44.8 & -29 43 13.0 & -102.0 & 2.58 & 24.2 & 7.2\\
2b  & HCN     & 17 43 44.8 & -29 43 13.0 & -107.0 & 2.08 & 20.5 & 5.2\\
2c\tablenotemark{1}  & HCO$^+$ & 17 43 44.8 & -29 43 16.0 & -103.9 & 2.51 & 26.9 & 4.8\\
2c  & HCN     & 17 43 42.7 & -29 42 58.0 & -92.1  & 1.24 & 23.0 & 11.6\\
2d  & HCO$^+$ & 17 43 43.2 & -29 43 25.0 & -105.6 & 1.90 & 25.4 & 9.5\\
2d  & HCN     & 17 43 42.9 & -29 43 22.0 & -102.0 & 2.31 & 23.5 & 11.8\\
3   & HCO$^+$ & 17 43 47.5 & -29 41 37.9 & -117.1 & 1.95 & 21.3 & 6.3\\
3   & HCN     & 17 43 48.0 & -29 41 34.0 & -116.9 & 2.27 & 22.1 & 9.7\\
3   & HCN     & 17 43 47.5 & -29 41 25.0 & -111.9 & 1.89 & 18.9 & 8.8\\ 
4a  & HCN     & 17 43 40.9 & -29 43 40.0 & -123.5 & 1.53 & 17.3 & 11.0\\
4a  & HCN     & 17 43 41.1 & -29 43 40.0 & -128.4 & 1.20 & 14.6 & 6.5\\
4b  & HCN     & 17 43 38.3 & -29 43 55.0 & -121.8 & 1.40 & 16.4 & 5.2\\
4b  & HCN     & 17 43 38.3 & -29 43 52.0 & -133.4 & 1.33 & 15.8 & 8.1\\
4b  & HCN     & 17 43 38.3 & -29 43 55.0 & -128.4 & 1.30 & 15.7 & 3.6\\
5   & HCN     & 17 43 45.9 & -29 45 10.0 & -178.0 & 1.24 & 20.0 & 9.6\\
6   & HCO$^+$ & 17 43 45.7 & -29 46 16.0 & -125.3 & 0.92 & 9.7 & 5.9\\
6   & HCN     & 17 43 45.9 & -29 46 13.0 & -126.8 & 1.28 & 18.6 & 5.8\\
6   & HCN     & 17 43 45.9 & -29 46 07.0 & -121.8 & 1.03 & 12.5 & 2.7\\
\hline \hline
7   & HCN     & 17 43 50.5 & -29 42 04.0 & -186.3 & 1.50 & 15.8 & 3.0\\
8   & HCN     & 17 43 52.1 & -29 41 19.0 & -103.7 & 1.48 & 17.9 & 13.8\\
9   & HCO$^+$ & 17 43 46.2 & -29 42 28.0 & -123.6 & 1.41 & 12.9 & 3.4\\ 
9   & HCN     & 17 43 45.7 & -29 42 28.0 & -120.2 & 1.29 & 13.4 & 5.1\\
10  & HCO$^+$ & 17 43 50.5 & -29 43 13.0 & -130.2 & 1.01 & 11.3 & 2.1\\
10  & HCN     & 17 43 51.0 & -29 43 31.0 & -115.2 & 1.11 & 13.0 & 3.5\\
11a & HCN     & 17 43 55.6 & -29 43 43.0 & -182.9 & 1.06 & 13.1 & 3.7\\
11b & HCN     & 17 43 55.1 & -29 44 34.0 & -184.6 & 0.85 & 16.3 & 7.6\\
12  & HCO$^+$ & 17 43 57.9 & -29 43 07.0 & -126.9 & 1.22 & 14.8 & 1.8\\
12  & HCN     & 17 43 58.1 & -29 43 13.0 & -123.49 & 0.87 & 13.0 & 5.6\\
13  & HCN     & 17 43 57.7 & -29 43 10.0 & -168.1 & 0.99 & 10.4 & 4.3\\
14  & HCN     & 17 43 50.8 & -29 44 34.0 & -115.2 & 0.80 & 11.7 & 3.1\\
15  & HCO$^+$ & 17 43 56.3 & -29 44 58.0 & -130.2 & 1.05 & 17.1 & 3.2\\
16\tablenotemark{2}  & HCN     & 17 43 40.2 & -29 41 28.0 & -105.3 & 2.17 & 11.1 & 4.1\\
16\tablenotemark{2}  & HCN     & 17 43 41.1 & -29 41 28.0 & -107.0 & 1.82 & 12.3 & 4.9\\
16\tablenotemark{2}  & HCN     & 17 43 41.6 & -29 41 19.0 & -111.9 & 1.38 & 9.9 & 7.1\\
16\tablenotemark{2}  & HCN     & 17 43 41.3 & -29 41 22.0 & -97.1  & 1.30 & 9.1 & 6.0\\
16\tablenotemark{2}  & HCN     & 17 43 41.3 & -29 41 37.0 & -115.2 & 1.20 & 16.3 & 10.5\\
17  & HCN     & 17 43 45.2 & -29 43 01.0 & -181.3 & 1.33 & 16.1 & 6.8\\
17  & HCN     & 17 43 45.2 & -29 42 55.0 & -188.0 & 1.26 & 9.9 & 4.4\\
17  & HCN     & 17 43 45.2 & -29 42 58.0 & -191.2 & 1.23 & 13.7 & 2.7\\
18  & HCN     & 17 43 50.1 & -29 42 13.0 & -97.1  & 0.93 & 12.5 & 3.3\\
18  & HCN     & 17 43 49.8 & -29 42 25.0 & -88.8  & 0.90 & 10.3 & 3.3\\
19  & HCN     & 17 43 51.0 & -29 43 31.0 & -115.2 & 1.11 & 13.0 & 3.5\\
\enddata

\tablenotetext{1}{The HCO$^+$ {\tt clumpfind} run connected 2b and 2c,
  whereas the HCN did not.}
\tablenotetext{2}{Clump 16 was halfway outside our criterion for
  rejection in the map edge; the rms error on the peak flux is greater
  than the other clumps, hence more clumps were found with the algorithm.}
\tablecomments{{\tt clumpfind} results for our data cube.  Clumps 1-6
  we consider ``resolved associations'' or ``clouds'' whereas 7-19 are
  ``unresolved clumps'' based on the 2$\sigma$ contours in the
  integrated map.  All features are found within $3^{\prime}$ of the
  map center. \label{measuredtable}}

\end{deluxetable}


\begin{deluxetable}{lccccc}
\tablenum{3}
\tabletypesize{\scriptsize}
\tablecaption{CARMA HCN Derived Properties}
\tablewidth{0pt}
\tablehead{
\colhead{Label} & \colhead{$\Delta v_{FWHM}$} & \colhead{$R$} &
\colhead{$M_{vir}$} & \colhead{$\log_{10}[\bar{n}(\text{H}_2)]$} &
\colhead{$t_{\text{cross}}$} \\ & 
 \colhead{(km s$^{-1}$)} & \colhead{(pc)} & \colhead{($10^4 M_{\odot}$)} &
 \colhead{log[(cm$^{-3}$)]} & \colhead{($10^5$ yr)}
}

\startdata

1   & 4.1  & 0.99 & 0.39 & 4.43 & 2.37 \\
2a  & 10.6 & 1.03 & 2.67 & 5.23 & 0.95 \\
2b  & 7.2  & 1.00 & 1.19 & 4.93 & 1.36 \\
2b  & 5.2  & 0.84 & 0.53 & 4.79 & 1.59 \\
2c  & 11.6 & 0.95 & 2.95 & 5.38 & 0.80 \\
2d  & 11.8 & 0.97 & 3.12 & 5.38 & 0.80 \\
3   & 9.7  & 0.91 & 1.98 & 5.26 & 0.92 \\
3   & 8.8  & 0.78 & 1.39 & 5.31 & 0.87 \\
4a  & 11.  & 0.71 & 1.99 & 5.59 & 0.63 \\
4a  & 6.5  & 0.60 & 0.59 & 5.28 & 0.91 \\
4b  & 5.2  & 0.67 & 0.42 & 4.98 & 1.27 \\
4b  & 8.1  & 0.65 & 0.99 & 5.40 & 0.79 \\
4b  & 3.6  & 0.65 & 0.19 & 4.70 & 1.76 \\
5   & 9.6  & 0.82 & 1.76 & 5.34 & 0.84 \\
6   & 5.8  & 0.77 & 0.60 & 4.97 & 1.29 \\
6   & 2.7  & 0.51 & 0.09 & 4.65 & 1.87 \\

\enddata


\tablecomments{\label{derivedtable} The clump radius in pc uses a
  galactocentric distance of 8.5 kpc.  When multiple clumps were
  detected in one resolved association, all clumps were included even
  if they were individually unresolved; we only calculate masses for
  resolved clouds.}
\end{deluxetable}


\begin{thebibliography}{}
\bibitem[Bally et al.(1987)]{bal87} Bally, J., Stark, A., Wilson, R.,
  Henkel, C. 1987, \apjs, 65, 13B
\bibitem[Bally \& Leventhal(1991)]{bal91} Bally, J. \& Leventhal,
  M. 1991, \nat, 353, 234B
\bibitem[Bania(1977)]{bania77} Bania, T.M. 1977, \apj, 216, 381
\bibitem[Binney et al.(1991)]{bin91} Binney, J. Gerhard, O., Stark,
  A., Bally, J. \& Uchida, K. 1991, \mnras, 252, 210
\bibitem[Bitran et al.(1997)]{bitran97} Bitran, M., Alvarez, H.,
  Bronfman, L., May, J. \& Thaddeus, P. 1997, A\&AS, 125, 99B
\bibitem[Ferri\`{e}re et al.(2007)]{fer07} Ferri\`{e}re, K., Gillard,
  W. \& Jean, P. 2007, \aap, 467, 611
\bibitem[Fux(1999)]{fux99} Fux, R. 1999, \aap, 345, 787F
\bibitem[Harris et al.(1994)]{harris94} Harris, A.I, Schuster, K.F.,
  Genzel, R., Plathner, B. \& Gundlach K.H. 1994, IJIMW, 15, 1465H
\bibitem[Harris et al.(1995)]{harris95} Harris, A.I., Avery, L.W.,
  Schuster, K.F., Tacconi, L.J. \& Genzel, R. 1995, \apj, 446L, 85H
\bibitem[Jenkins \& Binney(1994)]{jen94} Jenkins, A. \& Binney,
  J. 1994, \mnras, 270, 703
\bibitem[Larson(1981)]{larson81} Larson, R.B. 1981, \mnras, 194, 809L
\bibitem[Lee(1996)]{lee96} Lee, C.W. 1996, \apjs, 105, 129
\bibitem[Lepp \& Dalgarno(1996)]{lepp96} Lepp, S. \& Dalgarno,
  A. 1996, \aap, 306, L21
\bibitem[Linke et al.(1981)]{linke81} Linke, R., Stark., A, Frerking,
  M. 1981, \apj, 243, 147L
\bibitem[Liszt \& Burton(1978)]{liszt78} Liszt, H.S. \& Burton,
  W.B. 1978, \apj, 225, 815
\bibitem[Main et al.(1999)]{main99} Main, D.S., Smith, D.M., Heindl,
  W.A., Swank, J., Leventhal, M., Mirabel, I.F. \& Rodr\'{i}guez,
  L.F. 1999, \apj, 525, 901
\bibitem[Martin et al.(2004)]{mart04} Martin, C., Walsh, W., Xiao,
  K., Lane, A., Walker, C. \& Stark, A. 2004, \apjs, 150, 239
\bibitem[Mirabel et al.(1991)]{mir91} Mirabel, I., Morris, M., Wink,
  J., Paul, J. \& Cordier, B. 1991, \aap, 251, L43
\bibitem[Miyazaki \& Tsuboi(2000)]{miyazaki00} Miyazaki, A. \&
  Tsuboi, M. 2000, \apj, 536, 357M
\bibitem[Oka et al.(1998)]{oka98} Oka, T., Hasegawa, T., Sato, F.,
  Tsuboi, M., Miyazaki, A. 1998, \apjs, 118, 455O
\bibitem[Phillips \& Lazio(1995)]{phil95} Phillips, J. \& Lazio,
  T. 1995, \apj, 442, L37
\bibitem[Rodriguez-Fernandez et al.(2006)]{rod06}
  Rodriguez-Fernandez, N.J., Combes, F., Martin-Pintado, J.,
  Wilson, T.L. \& Apponi, A. 2006, \aap, 455, 963
\bibitem[Rodriguez-Fernandez \& Combes(2008)]{rod08} 
  Rodriguez-Fernandez, N.J. \& Combes, F. 2008, \aap, 489, 115 
\bibitem[Qin et al.(2008)]{qin08} Qin, S., Zhao, J., Moran, J.,
  Marrone, D., Patel, N., Wang, J., Liu, S., \& Kuan, Y. 2008, \apj,
  677, 353
\bibitem[Sault et al.(1995)]{sault95} Sault, R.J., Teuben, P.J. \&
  Wright, M.C.H. 1995, ASPC, 77, 433S
\bibitem[Stark et al.(1988)]{stark88} Stark, A., Bally, J., Knapp,
  G. \& Wilson, R. 1988, LNP, 315, 303S
\bibitem[Stark et al.(2004)]{stark04} Stark, A., Martin, C., Walsh,
  W., Xiao, K., Lane, A. \& Walker, C. 2004, \apj, 614, L41
\bibitem[Tsuboi et al.(1999)]{tsuboi99} Tsuboi, M., Handa, Toshihiro
  \& Ukita, Nobuharu 1999, \apjs, 120, 1T
\bibitem[Weidenspointner et al.(2008)]{weid08} Weidenspointner, G.,
  Skinner, G., Jean, P., Kn\"{o}dlseder, J., von Ballmoos, P.,
  Bignami, G., Diehl, R., Strong, A., Cordier, B., Schanne, S. \&
  Winkler, C. 2008, \nat, 451, 159
\bibitem[Williams et al.(1994)]{williams94} Williams, J., de Geus,
  E. \& Blitz, L. 1994, \apj, 428, 693W

\end{thebibliography}
\end{document}